\documentclass[twocolumn,showpacs,preprintnumbers,amsmath,amssymb]{revtex4}
\usepackage{graphicx}
\usepackage[ansinew]{inputenc}
\usepackage{color}
\def\beq{\begin{equation}}
\def\eeq{\end{equation}}

\begin{document}

\bibliographystyle{naturemag}

\title{Quantum simulation of the Dirac equation}

\author{R. Gerritsma$^{1,2}$}
\author{G. Kirchmair$^{1,2}$}
\author{F. Z\"ahringer$^{1,2}$}
\author{E. Solano$^{3,4}$}
\author{R. Blatt$^{1,2}$}
\author{C. F. Roos$^{1,2}$}
\email{christian.roos@uibk.ac.at}

\affiliation{$^1$Institut f\"ur Quantenoptik und Quanteninformation,
\"Osterreichische Akademie der Wissenschaften, Otto-Hittmair-Platz
1, A-6020 Innsbruck, Austria\\
$^2$Institut f\"ur Experimentalphysik, Universit\"at Innsbruck, Technikerstr.~25, A-6020 Innsbruck, Austria\\
$^3$Departamento de Qu\'{\i}mica F\'{\i}sica, Universidad del Pa\'{\i}s Vasco - Euskal Herriko Unibertsitatea, Apartado 644, 48080 Bilbao, Spain\\
$^4$IKERBASQUE, Basque Foundation for Science, Alameda Urquijo 36, 48011 Bilbao, Spain
}

\date{\today}

\begin{abstract}
The Dirac equation is a cornerstone in the history of physics \cite{Thaller:1992}, merging successfully quantum mechanics with special relativity, providing a natural description of the electron spin and predicting the existence of anti-matter \cite{Anderson:1933}. Furthermore, it is able to reproduce accurately the spectrum of the hydrogen atom and its realm, relativistic quantum mechanics, is considered as the natural transition to quantum field theory. However, the Dirac equation also predicts some peculiar effects such as Klein's paradox \cite{Klein:1929} and {\it Zitterbewegung}, an unexpected quivering motion of a free relativistic quantum particle first examined by Schr{\"o}dinger \cite{Schroedinger:1930}. These and other predictions would be difficult to observe in real particles, while constituting key fundamental examples to understand relativistic quantum effects. Recent years have seen an increased interest in simulations of relativistic quantum effects in different physical setups \cite{Garay:2000,Alsing:2005,Schliemann:2005,Lamata:2007,Bermudez:2007,Zhang:2008a,Vaishnav:2008,Otterbach:2009}, where parameter tunability allows accessibility to different physical regimes. Here, we perform a proof-of-principle quantum simulation of the one-dimensional Dirac equation using a single trapped ion \cite{Lamata:2007}, which is set to behave as a free relativistic quantum particle. We measure as a function of time the particle position and study {\it Zitterbewegung} for different initial superpositions of positive and negative energy spinor states, as well as the cross-over from relativistic to nonrelativistic dynamics. The high level of control of trapped-ion experimental parameters makes it possible to simulate elegant textbook examples of relativistic quantum physics.
\end{abstract}


\maketitle

The Dirac equation for a spin-1/2 particle with rest mass $m$ is given by:

\beq
i\hbar\frac{\partial \psi}{\partial t}=(c \,\mathbf{\alpha} \cdot \mathbf{\hat{p}}+\beta mc^2)\psi.
\eeq

Here $c$ is the speed of light, $\mathbf{\hat{p}}$ is the momentum operator, $\mathbf{\alpha}$ and $\beta$ are the Dirac matrices, which are usually given in terms of the Pauli matrices \cite{Thaller:1992}, and the wavefunctions $\psi$ are 4-component spinors. A general Dirac spinor $\psi$ can be decomposed into parts with positive and negative energies $E=\pm\sqrt{p^2c^2+m^2c^4}$. {\it Zitterbewegung} is understood as an interference effect between the positive and negative energy parts of the spinor and does not appear for spinors that consist consist entirely out of positive (or negative) energy. Furthermore, it is only present when these parts have significant overlap in position and momentum space and is therefore not a sustained effect under most circumstances \cite{Thaller:1992}. For a free electron, the Dirac equation predicts the {\it Zitterbewegung} to have an amplitude of the order of the Compton wavelength $R_{ZB} \sim 10^{-12}$~m and a frequency of $\omega_{ZB}\sim 10^{21}$~Hz and has so far been experimentally inaccessible. The real existence of {\it Zitterbewegung}, in relativistic quantum mechanics and in quantum field theory, has been a recurrent subject of discussion in the last years \cite{Krekora:2004,Wang:2008}.

Quantum simulation aims at simulating a quantum system with a controllable laboratory system underlying the same mathematical model. In this way, it is possible to simulate quantum systems that can neither be efficiently simulated on a classical computer \cite{Feynman:1982} nor easily accessed experimentally, while allowing parameter tunability over a wide range. The difficulties in observing real quantum relativistic effects have sparked significant interest in performing quantum simulations of their dynamics. Examples include black holes in Bose-Einstein condensates \cite{Garay:2000}, the Unruh effect in trapped ions \cite{Alsing:2005}, and {\it Zitterbewegung} for massive fermions in solid state physics \cite{Schliemann:2005}. Graphene is studied widely in connection to the Dirac equation \cite{Cserti:2006,Katsnelson:2006,Neto:2009}.

Trapped ions are particularly interesting for the purpose of quantum simulation \cite{Leibfried:2002,Porras:2004,Johanning:2009}, as they allow for exceptional control of experimental parameters, and initialization and read-out can be achieved with high fidelity. Recently, for example, a proof-of-principle simulation of a quantum magnet has been performed \cite{Friedenauer:2008} using trapped ions. The full three-dimensional Dirac equation Hamiltonian can be simulated with lasers coupling to the three vibrational eigenmodes and the internal states of a single trapped ion \cite{Lamata:2007}. The setup can be significantly simplified when simulating the Dirac equation in 1+1 dimensions, while the most stunning features of the Dirac equation remain, such as {\it Zitterbewegung} and the Klein paradox. In the Dirac equation in 1+1 dimensions,
\beq\label{eqDirac1g}
i\hbar\frac{\partial \psi}{\partial t}=H_D\,\psi=(c \,\hat{p}\,\sigma_x+mc^2\,\sigma_z)\psi,
\eeq
there is only one motional degree of freedom and the spinor is encoded in two internal levels, related to positive and negative energy states \cite{Lamata:2007}. For the velocity of the free Dirac particle we find $d\hat{x}/dt=[\hat{x},H_D]/i\hbar=c\sigma_x$ in the Heisenberg picture. For a massless particle $[\sigma_x,H_D]=0$ and hence $\sigma_x$ is a constant of motion. For a massive particle $[\sigma_x,H_D]\neq0$, and the evolution of the particle is described by

\beq
\hat{x}(t)=\hat{x}(0)+\hat{p}c^2H_D^{-1}t+i\,\hat{\xi}(e^{2iH_Dt/\hbar}-1),
\eeq

with $\hat{\xi}=\frac{1}{2}\hbar c(\sigma_x-\hat{p}cH_D^{-1})\, H_D^{-1}$. The first two terms represent an evolution linear in time, as expected for a free particle, while the last oscillating term may induce {\it Zitterbewegung}.

For the simulation, a single $^{40}$Ca$^+$ ion is trapped in a linear Paul trap\cite{Kirchmair:2009} with trapping frequencies $\omega_{ax}=2\pi\times$1.36~MHz axially and $\omega_{rad}=2\pi\times$3~MHz in the radial directions. Doppler cooling, optical pumping, and resolved sideband cooling on the $S_{1/2}\leftrightarrow D_{5/2}$ transition  in a magnetic field of 4~G prepare the ion in the axial motional ground state and in the internal state $|S_{1/2},m=1/2\rangle$. A narrow linewidth laser at 729~nm couples the states $|1\rangle := |S_{1/2},m=1/2\rangle$ and $|0\rangle := |D_{5/2},m=3/2\rangle$ which we identify as our spinor states. A bichromatic light field resonant with the upper and lower axial motional sidebands of the $|0\rangle\leftrightarrow|1\rangle$ transition with appropriately set phases and frequency realizes the Hamiltonian

\beq\label{eqDirac1}
H_D = 2 \eta \Delta \tilde{\Omega} \sigma_x \hat{p}+\hbar \Omega \sigma_z.
\eeq

Here, $\Delta =\sqrt{\hbar/2\tilde{m}\omega_{ax}}$ is the size of the ground-state wave function with $\tilde{m}$ the ion's mass, not to be confused with the mass $m$ of the simulated particle, $\eta=0.06$ is the Lamb-Dicke parameter, $\hat{p}=\frac{a^{\dag}-a}{2i}\frac{\hbar}{\Delta}$ is the momentum operator and $a^{\dag}$ and $a$ are the usual raising and lowering operators for the motional state along the axial direction. The first term in Eq.~\ref{eqDirac1} describes a state-dependent motional excitation with coupling strength $\eta\tilde{\Omega}$, corresponding to a displacement of the ion's wave packet in the harmonic trap. The second term is equivalent to an optical Stark shift and arises when the bichromatic light field is detuned from resonance by $2\Omega$. Equation~(\ref{eqDirac1}) reduces to the 1+1 dimensional Dirac Hamiltonian if we make the identifications $c:=2\eta\tilde{\Omega}\Delta$ and $mc^2:=\hbar\Omega$. The momentum and position of the Dirac particle are then mapped onto the corresponding quadratures of the trapped ion harmonic oscillator.

In order to study relativistic effects such as {\it Zitterbewegung} it is necessary to measure the expectation value $\langle \hat{x}(t)\rangle$ of the harmonic oscillator. It has been noted theoretically that such expectation values could be measured using very short probe times, without reconstructing the full quantum state \cite{Lougovski:2006,Santos:2007,Lamata:2007}. The unitary operator $U = exp(-i \eta \Omega_p \sigma_x \hat{x} t/\Delta)$, with $\hat{x}=(a^{\dag}+a)\Delta$ and probe Rabi frequency $\Omega_p$, transforms the observable $\sigma_z$ such that its slope for short probe times is proportional to $\langle \hat{x}\rangle$ (see methods). To measure $\langle \hat{x}\rangle$ we have to prepare the internal state in an eigenstate of $\sigma_y$, apply the unitary operator $U$ for short times and record the changing excitation. Since the Dirac Hamiltonian generally entangles the motional and internal state of the ion, we first trace over the internal state (see methods). Next, we prepare an eigenstate of $\sigma_y$ and apply the Hamiltonian generating $U$ for up to 14~$\mu$s, with 1-2~$\mu$s steps, with the probe Rabi frequency set to $\Omega_p=2\pi\times$~13~kHz. The slopes were obtained by a linear fit each based on $10^4$ to $3\cdot 10^4$ measurements.

\begin{figure}[t]
\includegraphics[width=8cm]{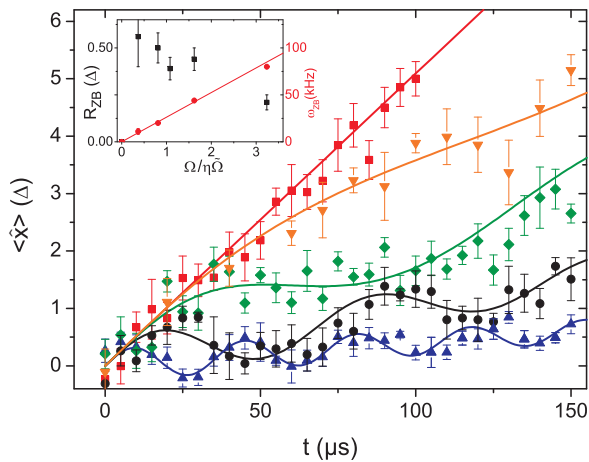}
\caption{\label{fig1:Zittercurves} Expectation values $\langle \hat{x}(t) \rangle$ for particles with different masses. The linear curve ($\textcolor[rgb]{1.00,0.00,0.00}{\blacksquare}$) represents a massless particle ($\Omega = 0$) moving with the speed of light given by $c=2\eta\tilde{\Omega}\Delta=0.052~\Delta/\mu$s for all curves. The other curves are for particles with increasing mass moving down from the linear curve. Their Compton wavelengths are given by $\lambda_C:=2\eta\tilde{\Omega}\Delta/\Omega=5.4\Delta$ ($\textcolor[rgb]{1,0.5,0.00}{\blacktriangledown}$), 2.5$\Delta$ ($\textcolor[rgb]{0.25,0.5,0.00}{\blacklozenge}$), 1.2$\Delta$ ($\bullet$) and 0.6$\Delta$ ($\textcolor[rgb]{0.0,0.00,1.00}{\blacktriangle}$), respectively. The solid curves represent numerical simulations. The figure shows {\it Zitterbewegung} for the crossover from the relativistic $2\eta\tilde{\Omega}\gg\Omega$ to the nonrelativistic limit $2\eta\tilde{\Omega}\ll\Omega$. The error bars are obtained from a linear fit assuming quantum projection noise. The inset shows fitted {\it Zitterbewegung} amplitude $R_{ZB}$ (${\blacksquare}$) and frequency $\omega_{ZB}$ ($\textcolor[rgb]{1.00,0.00,0.00}{\bullet}$) versus the parameter $\Omega/\eta\tilde{\Omega}$ (which is proportional to the mass). Error bars 1$\sigma$.}
\end{figure}

We simulate the Dirac equation by applying $H_D$ for varying amounts of time and for different particle masses. In the experiment we set $\tilde{\Omega}=2\pi \times 68$~kHz, corresponding to a simulated speed of light $c=$~0.052~$\Delta/\mu$s. The measured expectation values $\langle \hat{x}(t) \rangle$ are shown in Fig.~\ref{fig1:Zittercurves} for a particle initially prepared in the spinor state $\psi(x;t=0)= (\sqrt{2\pi}2\Delta)^{-\frac{1}{2}}e^{-\frac{x^2}{4\Delta^2}}\,\binom{1}{1}$ by sideband cooling and application of a $\pi/2$-pulse. {\it Zitterbewegung} appears for particles with non-zero mass, obtained by varying $\Omega$ in the range of $0<\Omega\le 2\pi\times$~13~kHz by changing the detuning of the bichromatic lasers.

We investigate the particle dynamics in the cross-over from relativistic to nonrelativistic dynamics. The data in Fig.~\ref{fig1:Zittercurves} were fitted with a model function of the form $\langle \hat{x}(t) \rangle = at + R_{ZB} \sin \omega_{ZB} t$. The fitted amplitude $R_{ZB}$ and frequency $\omega_{ZB}$ of the {\it Zitterbewegung} are shown in the inset of Fig.~\ref{fig1:Zittercurves}. From these data it can be seen that the frequency grows linearly with increasing mass  $\omega_{ZB} \approx 2\Omega$, whereas the amplitude decreases as the mass is increased. Since the mass of the particle is increased but the momentum and the simulated speed of light remain constant, the data in Fig.~\ref{fig1:Zittercurves} show the crossover from the far relativistic to nonrelativistic limits. Hence, the data confirm there is vanishing {\it Zitterbewegung} in both limits, as theoretically expected. In the far relativistic case this is because $\omega_{ZB}$ goes to zero, and in the nonrelativistic case because $R_{ZB}$ vanishes.
\begin{figure}
\includegraphics[width=8cm]{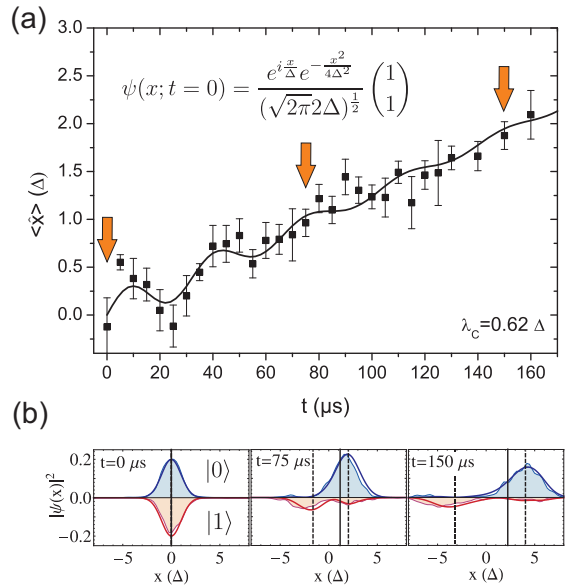}
\caption{\label{fig2:Otherinputs} {\it Zitterbewegung} for a state with non-zero average momentum. (a) Initially, {\it Zitterbewegung} appears due to interference of positive and negative energy parts of the state. As these parts separate, the oscillatory motion fades away. The solid curve represents a numerical simulation. (b) Measured (filled areas) and numerically calculated (solid lines) probability distributions $|\psi(x)|^2$  at the times $t=0$,~75 and 150~$\mu$s (as indicated by the arrows in (a)). The probability distribution corresponding to the state $|1\rangle$ is inverted for clarity. The vertical solid line represents $\langle \hat{x} \rangle$ as plotted in (a). The two dashed lines are the expectation values for the positive and negative energy parts of the spinor. Error bars 1$\sigma$.
}
\end{figure}

\begin{figure}
\includegraphics[width=8cm]{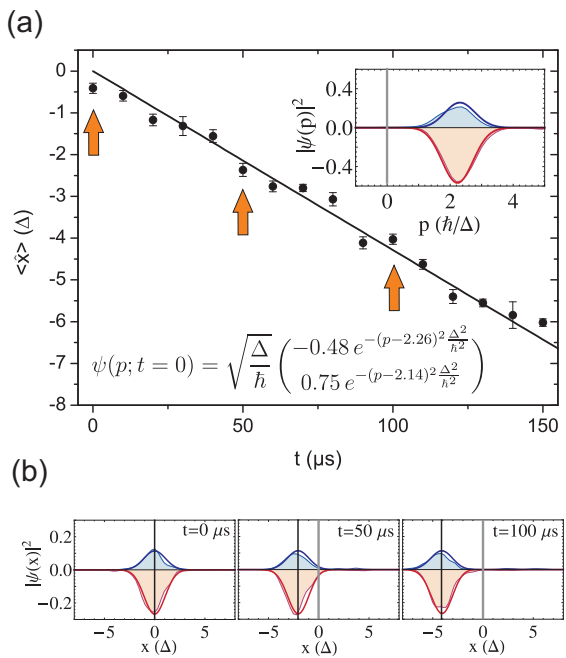}
\caption{\label{fig3:Negativeenergy} Time evolution for a negative energy eigenstate with $\lambda_C=1.2\Delta$. Laser pulses create a negative energy spinor with average momentum $\langle \hat{p} \rangle=2.2\hbar/\Delta$. The corresponding initial momentum distribution $|\tilde{\psi}(p)|^2$ is shown in the inset. The filled curves are data, the solid lines represent a numerical calculation. The curve in (a) shows no {\it Zitterbewegung}. The solid curve represents a numerical simulation. In (b) the measured probability distributions $|\psi(x)|^2$ for three different evolution times are shown (as indicated by the arrows in (a)). There is no splitting of the wavefunction and the evolution and spreading is as intuitively expected for a free particle. }
\end{figure}

The tools used for simulating the Dirac equation can also be applied to precisely set the initial state of the simulated particle. The particle in Fig.~\ref{fig2:Otherinputs}a was given an average initial momentum $\langle \hat{p}(t=0) \rangle=\hbar/\Delta$ by a displacement operation with the Hamiltonian $H=\hbar\eta\tilde{\Omega}\sigma_x\hat{x}/\Delta$. The initial state of this particle is built up out of a positive energy component with positive velocity and a negative energy component with negative velocity\cite{Thaller:2004}. As long as these parts overlap, {\it Zitterbewegung} is observed which dies out as the parts separate. Further information is obtained by a complete reconstruction of the probability distribution $|\psi(x)|^2$ shown in Fig.~\ref{fig2:Otherinputs}b. It is also possible to initialize the spinor in a pure negative or positive energy spinor (see methods). In Fig.~\ref{fig3:Negativeenergy}a, the time evolution $\langle \hat{x}(t) \rangle$ of a negative energy spinor with average momentum $\langle \hat{p} \rangle=2.2\hbar/\Delta$ is shown. The corresponding reconstructed probability distributions are displayed in Fig.~\ref{fig3:Negativeenergy}b and it can be seen that there is indeed no {\it Zitterbewegung} or splitting of the wavefunction.

We have implemented a proof-of-principle quantum optical simulation of a tunable relativistic quantum mechanical system. We have demonstrated that the simulated one-dimensional Dirac dynamics for a free particle shows {\it Zitterbewegung} and several of its counterintuitive quantum relativistic features. A natural route for the near future will be to move theoretically and experimentally towards the simulation of dynamics that are impossible (or difficult) to calculate in real systems, such as in quantum chemistry \cite{Aspuru-Guzik:2005} or quantized Dirac fields in a quantum field theory context \cite{Thaller:1992}. We consider this experiment to be an important first step that will pave the way towards more complex quantum simulations. Furthermore, the mapping between quantum optical systems and relativistic quantum mechanics may be followed by further analogies between the Dirac dynamics and the Jaynes-Cummings model \cite{Rozmej:1999,Bermudez:2007,Bermudez:2007b} and photonic \cite{Zhang:2008a} or sonic analogies \cite{Zhang:2008b}.


\section*{Methods}
{\bf Measurement of $\mathbf{\langle x \rangle}$ and $\mathbf{|\psi(x)|^2}$.} In ion trap experiments the only observable that can directly be measured by fluorescence detection is $\sigma_z$. Additional laser pulses can be used to map other observables onto $\sigma_z$. In the experiment we apply a state-dependent displacement operation $U=\exp(-ik\hat{x}\sigma_x/2)$ to the quantum state $\rho$ followed by a measurement of $\sigma_z$ which is equivalent to measuring the observable
\beq
A(k)=U^\dagger\sigma_zU=\cos(k\hat{x})\sigma_z+\sin(k\hat{x})\sigma_y.
\eeq
Here, $k=2\eta\Omega_p t/\Delta$ is proportional to the interaction time $t$. If the ion's internal initial state is the eigenstate of $\sigma_z$ belonging to eigenvalue $+1$, $\langle A(k)\rangle=\langle\cos(k\hat{x})\rangle$ and for the eigenstate of $\sigma_y$ belonging to eigenvalue $+1$, $\langle A(k)\rangle=\langle\sin(k\hat{x})\rangle$. A Fourier transformation of these measurements yields the probability density $\langle\delta(\hat{x}-x)\rangle$ in position space.

For the position operator we have that $\left.\frac{d}{dk}\langle A(k)\rangle\right\vert_{t=0}\propto\langle\hat{x}\sigma_y\rangle$. Measuring $\langle \hat{x} \rangle$ thus requires the preparation of an eigenstate of $\sigma_y$ which however cannot be done directly when the motional state is entangled with the internal state. To solve this problem we first incoherently recombine the internal state in $|1\rangle$. This is done by first shelving the population initially in $|1\rangle$ to $|D_{5/2},m=5/2\rangle$ using a Rapid Adiabatic Passage transfer (RAP). A second RAP transfers the population in $|0\rangle$ to $|1\rangle$. A 100~$\mu$s laser pulse at 854~nm transfers the population in $|D_{5/2},m=5/2\rangle$ to $|P_{3/2},m=3/2\rangle$ from which it spontaneously decays to $|1\rangle$. The transfer efficiency is $>99$~\%, limited by the small branching ratio to the $D_{3/2}$ state. In the transfer steps a probability exists that the motional state of the ion is changed. This probability is however very small due to the small Lamb-Dicke parameter but could be eliminated completely by a separate measurement of the motional states of the spinor states $|1\rangle$ and $|0\rangle$, at the expense of a longer data acquisition time.

To distinguish between populations in the state $|0\rangle$ and $|1\rangle$, when reconstructing $|\psi(x)|^2$ (as shown in Figs.~\ref{fig2:Otherinputs} and \ref{fig3:Negativeenergy} a short (200~$\mu$s) fluorescence detection was applied to measure the internal state. Only cases where $|0\rangle$ was measured (leaving the motional state unchanged as no photons were scattered) were used for the subsequent analysis. To reconstruct $|\psi(x)|^2$ belonging to $|1\rangle$, a $\pi$-pulse prior to the short detection was used to interchange the internal state populations. \newline

{\bf Constructing a pure negative energy spinor.} A general spinor is built up out of positive and negative energy components $E_\pm=\pm\sqrt{c^2p^2+m^2c^4}$ such that $\psi=P^+\psi+P^-\psi$. In momentum space, the projection operators are given by

\beq
P^{\pm}(p)=\frac{1}{2}\left(I_2\pm\frac{cp\sigma_x+mc^2\sigma_z}{\sqrt{c^2p^2+m^2c^4}}\right).
\eeq

Here, $I_2$ is the 2$\times$2 identity matrix. The spinor state in Fig.~\ref{fig3:Negativeenergy} was engineered backwards by projecting out the negative energy part of a wavepacket with average momentum $\langle \hat{p} \rangle=2.2\hbar/\Delta$ and renormalizing the spinor.

The complete sequence for approximating the negative energy state is conveniently described in the basis of the eigenstates $|\pm\rangle_y$ of $\sigma_y$. After ground state cooling we prepare the state $|+\rangle_y$, then we displace this state to an average momentum state $\langle \hat{p} \rangle=2.2\hbar/\Delta$ by the displacement Hamiltonian $H=\hbar\eta\tilde{\Omega}\sigma_y\hat{x}/\Delta$. Next, a far detuned laser pulse rotates the internal state to $0.84|+\rangle_y+i0.53|-\rangle_y$. The displacement Hamiltonian $H=-\hbar\eta\tilde{\Omega}\sigma_y\hat{x}/\Delta$ shifts these parts in opposite directions to create the required asymmetry between the average momenta of the components. A final $\pi/2$-pulse creates the state shown in Fig.~\ref{fig3:Negativeenergy}. This state has $>99$\% overlap with the desired negative energy state.


\section*{Acknowledgements}
We gratefully acknowledge support by the Austrian Science Fund (FWF), by the European Commission (Marie-Curie program) and by the Institut f\"ur Quanteninformation GmbH. E.S. acknowledges support of UPV-EHU Grant GIU07/40 and EU project EuroSQIP. This material is based upon work supported in part by IARPA. We thank Hartmut H{\"a}ffner and Mikhail Baranov for comments on the manuscript.


\end{document}